\documentclass[preprint,showpacs,preprintnumbers,superscriptaddress,amsmath,amssymbm,aps]{revtex4}
\usepackage{graphicx}
\usepackage{dcolumn}
\usepackage{bm}
\usepackage[american]{babel}

\begin{document}

\title{Experimental observation of optical rotation\\ generated in vacuum by a magnetic field}

\author{E.~Zavattini}
\affiliation{Istituto Nazionale di Fisica Nucleare (INFN), Sezione di Trieste and Universit\`a di Trieste, Trieste, Italy}
\author{G.~Zavattini}
\affiliation{Istituto Nazionale di Fisica Nucleare (INFN), Sezione di Ferrara and Universit\`a di Ferrara, Ferrara, Italy}
\author{G.~Ruoso}
\affiliation{Istituto Nazionale di Fisica Nucleare (INFN), Laboratori Nazionali di Legnaro, Legnaro, Italy}
\author{E.~Polacco}
\affiliation{Istituto Nazionale di Fisica Nucleare (INFN), Sezione di Pisa and Universit\`a di Pisa, Pisa, Italy}
\author{E.~Milotti}
\affiliation{Istituto Nazionale di Fisica Nucleare (INFN), Sezione Trieste and Universit\`a di Udine, Udine, Italy}
\author{M.~Karuza}
\affiliation{Istituto Nazionale di Fisica Nucleare (INFN), Sezione di Trieste and Universit\`a di Trieste, Trieste, Italy}
\author{U.~Gastaldi}
\affiliation{Istituto Nazionale di Fisica Nucleare (INFN), Laboratori Nazionali di Legnaro, Legnaro, Italy}
\author{G.~Di Domenico}
\affiliation{Istituto Nazionale di Fisica Nucleare (INFN), Sezione di Ferrara and Universit\`a di Ferrara, Ferrara, Italy}
\author{F.~Della Valle}
\affiliation{Istituto Nazionale di Fisica Nucleare (INFN), Sezione di Trieste and Universit\`a di Trieste, Trieste, Italy}
\author{R.~Cimino}
\affiliation{Istituto Nazionale di Fisica Nucleare (INFN), Laboratori Nazionali di Frascati, Frascati, Italy}
\author{S.~Carusotto}
\affiliation{Istituto Nazionale di Fisica Nucleare (INFN), Sezione di Pisa and Universit\`a di Pisa, Pisa, Italy}
\author{G.~Cantatore}
\email[corresponding author: ]{cantatore@trieste.infn.it}
\affiliation{Istituto Nazionale di Fisica Nucleare (INFN), Sezione di Trieste and Universit\`a di Trieste, Trieste, Italy}
\author{M.~Bregant}
\affiliation{Istituto Nazionale di Fisica Nucleare (INFN), Sezione di Trieste and Universit\`a di Trieste, Trieste, Italy}
\collaboration{PVLAS Collaboration}
\noaffiliation

\date{\today}
\begin{abstract}
We report the experimental observation of a light polarization rotation in vacuum in the presence of a transverse magnetic field. Assuming that data distribution is Gaussian, the average measured rotation is $(3.9\pm0.5)\;10^{-12}$~rad/pass, at 5~T with 44000 passes through a 1~m long magnet, with $\lambda$ = 1064 nm. The relevance of this result in terms of the existence of a light, neutral, spin-zero particle is discussed.
\end{abstract}


\maketitle

{\it Introduction} -- The existence of quantum fluctuations opens the possibility of considering vacuum itself as a material medium with dielectric properties. One such predicted property is the fact that vacuum becomes birefringent in the presence of a magnetic field due to photon-photon interactions \cite{QED,Adler,Iacopini}. Furthermore, the Primakoff effect, where light, neutral, scalar or pseudoscalar particles are produced from a two-photon vertex, could also give rise to macroscopically observable optical phenomena. In this second case it is expected that, by applying an external magnetic field, vacuum becomes birefringent and dichroic \cite{Anselm,Maiani,Raffelt}. The result of the induced dichroism is that the polarization plane of linearly polarized light is rotated by an angle 
\begin{equation}
\alpha=\frac{(a_{\parallel}-a_{\perp})}{2}\:D\:\sin2\theta
\end{equation}
after traveling a distance $D$ in a direction perpendicular to the field. Here $a_{\parallel}$ and $a_{\perp}$ are the absorption coefficients of the parallel and perpendicular components of the light polarization with respect to the magnetic field, and $\theta$ is the angle between the magnetic field and the polarization vector.

The PVLAS collaboration is studying the magnetic birefringence and dichroism of vacuum with a high-sensitivity ($\sim10^{-7}$~rad/$\sqrt{\mbox{Hz}}$) optical ellipsometer, at the INFN Legnaro National Laboratory, Legnaro, Italy \cite{Bakalov}. In this apparatus, a linearly polarized light beam at 1064~nm from a Nd:YAG laser interacts, inside a Fabry-Perot optical resonator, with the field produced by a superconducting, 1~m long dipole magnet  operated at a maximum field of 5.5~T.

In this letter we report the first observation of a rotation of the polarization plane induced in vacuum on linearly polarized light by a transverse magnetic field. This signal is generated within the Fabry-Perot cavity when the magnet is energized. Data from these measurements can be interpreted in terms of the existence of a very light, neutral, spin-zero particle coupled to two photons.

\begin{figure}
\includegraphics[width=12cm]{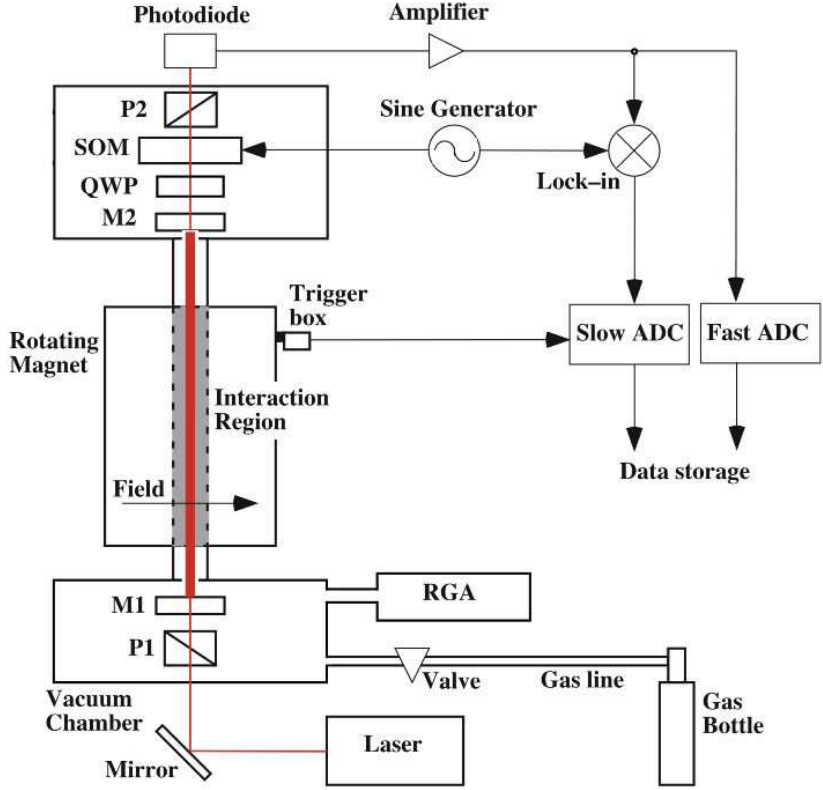}
\caption{Schematic layout of the PVLAS experimental apparatus. P1 and P2 crossed polarizing prisms, M1 and M2 Fabry-Perot cavity mirrors, QWP quarter-wave plate, SOM ellipticity modulator, RGA residual gas analyzer.}
\end{figure}

{\it Apparatus and experimental technique} -- Figure 1 shows a schematic layout of the PVLAS apparatus. The 1~m long interaction region is contained within a 4.6~m long, 25~mm diameter quartz tube. This tube is placed vertically and traverses the bore of the dipole magnet. The magnet is housed within a 5~ton, 3.1~m high, warm-bore cryostat, and the field direction lies in the horizontal plane: the cryostat itself sits on a 0.9~m radius turntable standing on a concrete beam fastened to the experimental hall floor \cite{Pengo}. The UHV vacuum chambers hosting the optical elements, placed at the top and at the bottom of the tube, are fastened to two granite optical benches. These benches are fixed on a concrete platform mechanically isolated from the rest of the hall floor. During normal operation the cryostat is filled with liquid He at 4.2~K and the magnet is energized with a current of about 2030~A, resulting in a 5.5~T field over the entire interaction region. To allow rotation of the magnet, the coils are shorted and disconnected from the power supply. The field intensity is monitored by a set of Hall probes \cite{Pengo}. The turntable is actuated by a low-vibration hydraulic drive and normally rotates the magnet-cryostat assembly, around a vertical axis, at a frequency $\nu_{m}\sim0.3$~Hz. During vacuum measurements, the quartz tube and the two vacuum chambers are kept in UHV conditions ($P\sim10^{-8}$~mbar) by a liquid N$_{2}$ trap combined with Ti sublimation pumps. This pumping scheme has been chosen in order to avoid mechanical vibrations and possible couplings between the rotating dipole field and the ion-pump permanent magnets. The residual gas composition is monitored by means of a residual gas analyzer (RGA). 
The interaction region is contained within a high-finesse Fabry-Perot optical resonator (FP) formed by a pair of dielectric, multilayer, high-reflectivity, 11~m curvature radius mirrors placed 6.4~m apart (M1 and M2). Resonator diffraction losses are found to be negligible. The ellipsometer consists of a pair of crossed polarizing prisms (P1 and P2) together with an ellipticity modulator (Stress Optic Modulator, or SOM \cite{Brandi}). The SOM, driven by a Sine Generator at a frequency $\nu_{\mbox{\scriptsize{SOM}}}=506$~Hz, provides a carrier ellipticity which beats with the modulated signal produced by magnet rotation and allows heterodyne detection. A quarter-wave plate (QWP), inserted between the M2 mirror and the SOM, can be placed in or removed from the beam path without breaking the vacuum. It can also be rotated in order to exchange its fast and slow axes. 
The laser beam (from a 100~mW, CW, Nd:YAG laser at 1064~nm) is frequency-locked to the FP cavity by means of an electro-optical feedback loop \cite{Cantatore}. In this way, the optical path is increased by a factor $N=2{\cal F}/\pi$, where ${\cal F}$ is the finesse of the FP. Typically, for the PVLAS cavity, ${\cal F}\sim7\;10^{4}$.
Light transmitted through the analyzer P2 is detected by a photodiode followed by a high gain ($10^{7}$~V/A) transimpedance amplifier, and the resulting voltage signal is fed to the digitizing electronics. The position of the rotating table is sensed by an optoelectronic position detector and this signal is also read and digitized: this allows determination of absolute phases.

{\it Measurements and results} -- The light intensity transmitted through the crossed polarizers of the PVLAS ellipsometer is \cite{Bakalov}
\begin{equation}
\label{intensity}
I=I_{0}\left\{\sigma^{2}+\left[\alpha(t)+\eta(t)+\Gamma(t)\right]^{2}\right\}
\end{equation}
where $I_{0}$ is the intensity before the analyzer P2, $\sigma^{2}$ is the extinction factor, $\alpha(t)~=~\alpha_{0}\:\cos(4\pi\nu_{m}t+2\theta_{m})$ is the rotation to be detected, $\eta(t)=\eta_{0}\:\cos(2\pi\nu_{\mbox{\scriptsize{SOM}}}t+\theta_{\mbox{\scriptsize{SOM}}})$ is the carrier ellipticity, and $\Gamma(t)$ represents additional, quasi-static, uncompensated rotations and ellipticities. Table I lists the Fourier components obtained by expanding expression \ref{intensity}, and it shows that the amplitude of the two sidebands separated by $2\nu_{m}$ (twice the magnet rotation frequency) from the carrier frequency at $\nu_{\mbox{\scriptsize{SOM}}}$ is proportional to the the amplitude of the rotation to be detected. When properly aligned, the QWP transforms apparent rotations (dichroisms) into ellipticities \cite{Born}, which then beat with the carrier ellipticity signal and are detected. These ellipticities conserve amplitude and undergo a sign change if the fast axis of the QWP is at 90$^{\circ}$ with respect to the initial polarization direction. We label this setting as ÒQWP $90^{\circ}$Ó, and use ÒQWP $0^{\circ}$Ó when the fast axis of the QWP is aligned with the initial polarization direction. In both cases, ellipticities are converted into rotations and pass undetected. Conversely, without the QWP in the beam path, ellipticities acquired in the interaction region beat with the SOM carrier ellipticity and are detected, while acquired rotations do not beat with the SOM, and pass undetected.

\begin{table}
\begin{tabular}{ccc}
\hline\hline
Frequency & Intensity/$I_{0}$ & Phase\\\hline
0 & $\sigma^{2}+\Gamma^{2}+\eta_{0}^{2}/2$+$\alpha_{0}^{2}/2$\\
$4\nu_{m}$ & $\alpha_{0}^{2}/2$ & $4\theta_{m}$\\
$\nu_{\mbox{\scriptsize{SOM}}}$ & $2\Gamma\eta_{0}$ & $\theta_{\mbox{\scriptsize{SOM}}}$\\
$\nu_{\mbox{\scriptsize{SOM}}}\pm2\nu_{m}$ & $\eta_{0}\alpha_{0}$ & $\theta_{\mbox{\scriptsize{SOM}}}\pm2\theta_{m}$\\
$2\nu_{\mbox{\scriptsize{SOM}}}$ & $\eta_{0}^{2}/2$ & $2\theta_{\mbox{\scriptsize{SOM}}}$\\
\hline\hline
\end{tabular}
\caption{Main frequency components of the detection photodiode signal.}
\end{table}

As a check of the Fourier phase of observed physical signals, we have measured the Cotton-Mouton effect (CME) \cite{Carusotto} in N$_{2}$ and Ne at several pressure values (for N$_{2}$ pressure ranged from 6~$\mu$bar to 1.8~mbar, and for Ne from 0.5~mbar to 20~mbar). For these measurements, the gas was inserted in the interaction region, and the QWP was momentarily removed. Sideband peaks were then observed in the photodiode signal Fourier spectrum at twice the magnet rotation frequency from the carrier. The phases of these peaks correspond to the angular position of the turntable at which the field direction is at $45^{\circ}$ with respect to the initial fixed polarization. In this position the magnetic birefringence due to the CME is maximum, and the numerical value of the phase is fixed by the geometry of the apparatus. Phase and amplitude data resulting from these measurements were plotted on a polar plane and fitted with radial lines. These lines fix what we call the physical axis, and are in agreement with geometrical considerations.

\begin{figure}
\includegraphics[width=12cm]{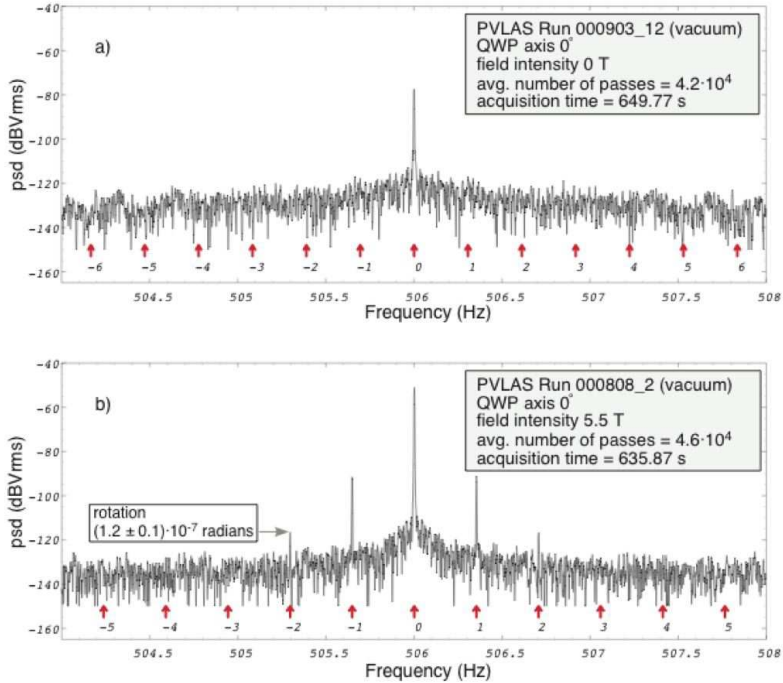}
\caption{Typical Fourier amplitude spectra from vacuum rotation data ($P\sim10^{-8}$~mbar) with and without magnetic field. Both spectra were taken with the magnet rotating. Arrows and numbers below the curves indicate the expected position of sidebands with frequency shifts that are integer multiples of $\nu_{m}$; the relevant signal peaks correspond to frequency shifts of $\pm 2\nu_{m}$.}
\end{figure}

\begin{figure}
\includegraphics[width=12cm]{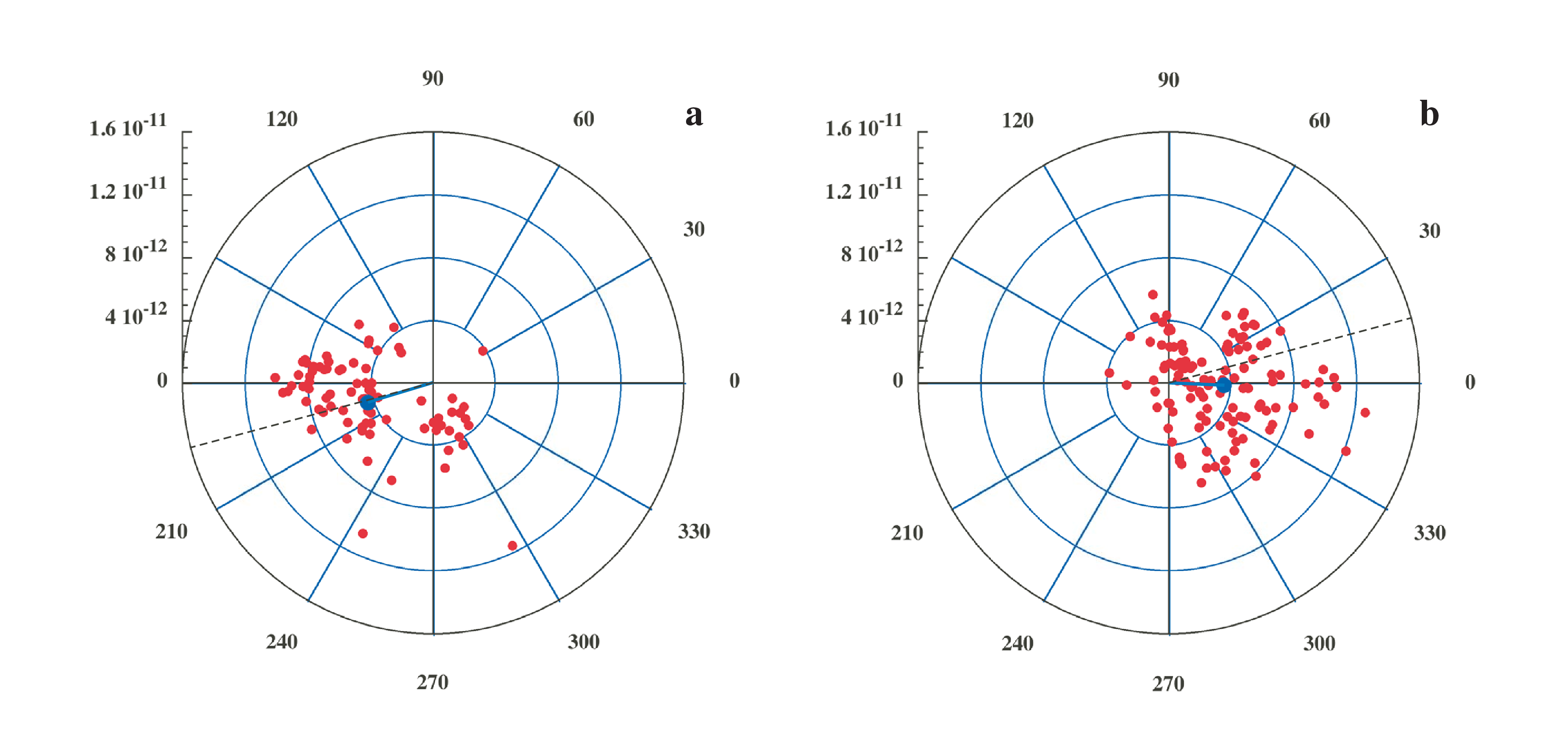}
\caption{Polar plot showing the distribution of rotation data in vacuum with $B=5$~T and (a) the QWP in the $90^{\circ}$ position, and (b) the QWP in the $0^{\circ}$ position. Each data point corresponds to a single 100 s record. The statistical uncertainties from the fit procedure are small ($\sim10^{-12}$~rad/pass) and are omitted for clarity. Solid lines represent the resulting average vectors and the dotted line the physical axis.}
\end{figure}

Two typical Fourier amplitude spectra of the detection photodiode signal in vacuum are shown in Figure 2. Figure 2a shows a spectrum around the 506 Hz carrier frequency when the magnet is off: this spectrum does not show sideband peaks at $\pm 2\nu_{m}$ from the carrier.  Figure 2b shows an amplitude spectrum observed in the same conditions of Figure 2a, but with the magnet on at 5.25~T. Here sideband peaks at $\pm 2\nu_{m}$ are present (peak phase is $336^{\circ}\pm6^{\circ}$). To simplify the comparison, we show spectra with a similar noise level ($\sim10^{-8}$~rad) near the ``2" sidebands. The sideband peaks at frequency shifts $\pm \nu_{m}$ were studied and found to be uncorrelated, both in amplitude and in phase, to the signal sidebands at $\pm 2\nu_{m}$. Sideband peaks at a distance $\pm 2\nu_{m}$ from the carrier, similar to the ones appearing in Figure 2b, have been observed in all the PVLAS data runs with the magnet energized, and
sideband peak phases have been measured in addition to their amplitudes. Several tests have been performed to clarify the nature of these peaks. We find that the peaks are present only when the magnet is on and the mirrors of the FP cavity are mounted. We have also observed that when QWP is removed from the beam the peak amplitude changes. Further, when the fast and the slow optical axes  of the QWP are exchanged, that is the QWP is rotated by $90^{\circ}$, signal phase changes by 180$^{\circ}$ as expected for a ``true'' rotation \cite{Born}, while the amplitude remains unchanged. These facts indicate that the ``$\pm 2\nu_m$'' sidebands are associated with an interaction inside the FP cavity.

Vacuum data have been collected for two years, in the course of six two-week runs, and were routinely taken with cavity finesses greater than 62000. Collected data were subdivided in 100~s long records, for a total of 84 records at QWP $90^{\circ}$ and 121 records at QWP $0^{\circ}$. Each record was then analyzed with a fitting procedure that compensates the mechanical jitter of the rotating table and the uneven sampling of the table position \cite{Milotti}. For each record, the procedure returns the $x$ and $y$ components of a vector in the amplitude-phase plane of the  ``$\pm 2\nu_m$'' sidebands; these components are Gaussian variates, and the procedure returns their statistical variances as well. Since these data records have been taken with different cavity finesses, to plot all available rotation data records in a single graph amplitudes have been normalized to a single pass in the cavity. The measurements performed with the two different QWP orientations, taken within the same two-week run, appear in the $x-y$ plane as two groups of points located on opposite sides with respect to the origin. However, the average phase and amplitude of each group vary from one run to the other more than their individual dispersion. As a consequence, the standard deviation of the entire set of $x$ and $y$ components shown in Figure 3 is larger than the standard deviation obtained from the fit procedure: the values are (a)  $(\sigma_{S,x}, \sigma_{S,y}) = (2.3, 1.8)~10^{-12}$~rad/pass, and  (b) $(\sigma_{S,x}, \sigma_{S,y}) = (2.4, 2.2)~10^{-12}$~rad/pass. We interpret this as an indication of the presence of unidentified systematic effects which increase the overall measurement uncertainty. Assuming that these systematic effects have a Gaussian distribution in the long term, we estimate additional variances $\sigma_{S,x}^2$, $\sigma_{S,y}^2$ from the distributions of the $x$ and $y$ components. Thus we associate to each $x$ and $y$ component a new variance $\sigma_{tot}^2$ which is the sum of the statistical variance from the fit procedure and of the additional variances $\sigma_{S,x}^2$ and $\sigma_{S,y}^2$, and calculate weighted averages of $x$ and $y$ for each orientation of the QWP. Figure 4 shows both the weighted average vectors, 0 and 90, and the half-difference vector $\Delta$, representing that part of the signal changing sign under a QWP axis exchange. The half-sum vector $\Sigma$ is a systematic error. This analysis suggests a physical origin for the rotation signal observed in vacuum when the magnetic field is present (numerical values for $\Sigma$ and $\Delta$ are listed in Table II).
From Table II we find the following value for the amplitude of the measured rotation in vacuum with $B \approx 5$~T (quoted with a $3\sigma$ uncertainty interval):

\begin{equation}
\label{alpha_exp}
\alpha=(3.9\pm0.5)\;10^{-12}\mbox{~rad/pass}
\end{equation}

\begin{figure}
\includegraphics[width=12cm]{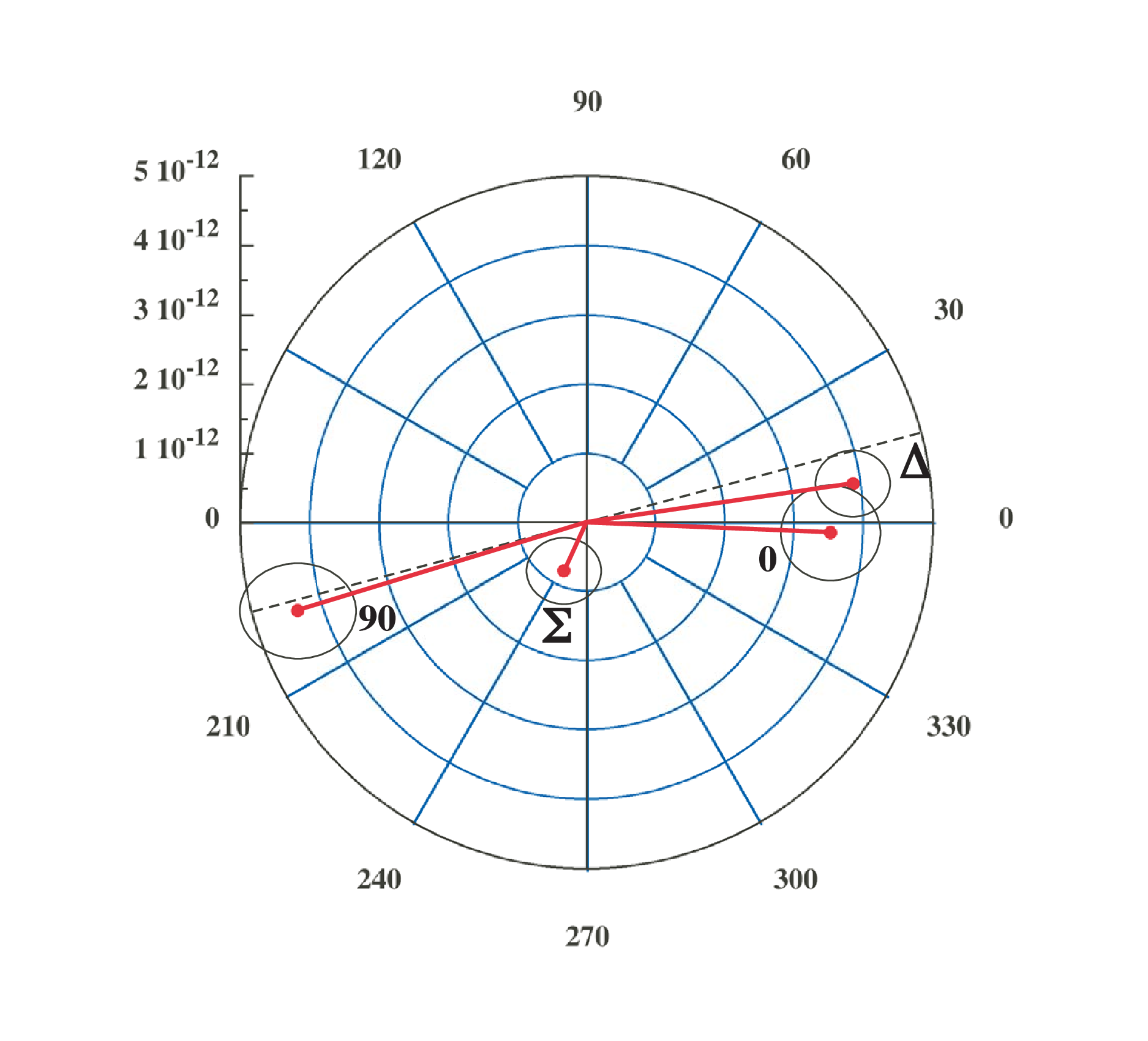}
\caption{Polar plot showing the weighted averages of the vectors shown in Figure 3. Angles are measured in degrees and amplitudes in rad/pass. The dotted line at $15.1^{\circ}$ shows the direction of the physical axis. The vectors $0$ and $90$ represent the average rotations for two orientations of the QWP, with ellipses at their tips giving the $3\sigma$ uncertainty regions. The vectors $\Delta$ and $\Sigma$ are the half-difference and  half-sum of the average vectors, respectively.}
\end{figure}

\begin{table}
\begin{tabular}{ccccccccc}
\hline\hline
& \multicolumn{2}{c}{$0^{\circ}$} &\multicolumn{2}{c}{$90^{\circ}$} & \multicolumn{2}{c}{$\Delta$} & \multicolumn{2}{c}{$\Sigma$}\\
& $x$ & $y$  & $x$ & $y$  & $x$ & $y$  & $x$ & $y$ \\
\hline
& 3.51 & -0.15 & -4.16 & -1.26 & 3.83 & 0.55 & -0.33 & -0.70\\
$\sigma_{stat}$ & 0.07 & 0.07 & 0.07 & 0.07 & \multicolumn{4}{c}{ }\\
$\sigma_{tot}$ & 0.24 & 0.23 & 0.28 & 0.23 & 0.18 & 0.16 & 0.18 & 0.16 \\
\hline\hline
\end{tabular}
\caption{Values and corresponding standard deviations, both statistical (stat) and total (tot), in units of $10^{-12}$~rad/pass, of the $x$ and $y$ components of the vectors shown in Fig. 4.}
\end{table}



{\it Discussion} -- We have observed a rotation of the polarization plane of light propagating through a transverse magnetic field having the following characteristics:

-it depends on the presence of the magnetic field;

- it is localized within the Fabry-Perot cavity;

-it has the proper phase with respect to the magnetic field instantaneous direction.

Experimentally, then, we find that vacuum induces a rotation of the polarization plane in the presence of a magnetic field.  The possibility that this effect is due to an unknown, albeit very subtle, instrumental artifact has been investigated at length without success.  A brief listing of the possible spurious effects which have been considered includes: electrical pick-ups, effects from the residual gas, beam movements induced by couplings of the rotating field to parts of the apparatus (the permanents magnets of the ion vacuum pumps, for instance), higher harmonic generation by the sideband peaks at $\pm \nu_{m}$, direct effect of the field on the optics.

Speculations on the physical origin of the measured signal suggest the possible presence of new physics. In fact, a selective absorption of photons, resulting in an apparent rotation, could be caused by photon splitting processes \cite{Adler}, where photons interact with the external field and split into two or more lower energy photons. These photons would not resonate with the Fabry-Perot cavity and would go undetected through the PVLAS ellipsometer, where this effect would be seen as an absorption. However, in our apparatus photon splitting is shown to be negligible.

Another source of apparent rotation could be associated to a neutral, light boson produced by a two-photon vertex \cite{Maiani,Raffelt,Mass˜}. In this photon-boson oscillation framework, both induced rotation and induced ellipticity can be expressed as functions of the particle mass $m_{b}$ and of the inverse coupling constant to two photons $M_{b}$ \cite{Nota_1}. In particular, the amplitude of the induced dichroism $\varepsilon$ is given by  (in natural Heaviside-Lorentz units)
\begin{equation}
\label{epsi_rotation}
\varepsilon=\sin2\theta\:\left(\frac{BL}{4M_{b}}\right)^{2}N\left[\frac{\sin\left(m_{b}^{2}L/4\omega\right)}{m_{b}^{2}L/4\omega}\right]^{2}
\end{equation}
where $\theta$ is the angle between the magnetic field $B$ and the light polarization direction, $L$ is the length of the magnetic region, $N$ is the number of passes in the FP cavity, and $\omega$ is the photon energy. Therefore, the measured rotation in vacuum fixes a relationship between mass $m_{b}$ and inverse coupling $M_{b}$.
It is important to note that if we set in formula \ref{epsi_rotation} the diffraction term $\left[\frac{\sin\left(m_{b}^{2}L/4\omega\right)}{m_{b}^{2}L/4\omega}\right]^{2}=1$, the constant $M_{b}$ implied by result \ref{alpha_exp} is bound by $M_{b} \lesssim 6\;10^{5}$~GeV. This is compatible with rotation limits published in \cite{Cameron}. When these limits are combined with the $M_{b}$~vs.~$m_{b}$ curve corresponding to the PVLAS signal, we find that only one segment of this curve is allowed as a region for possible $(m_{b},M_{b})$ pairs. The allowed segment is contained in the intervals $2\;10^{5}\mbox{~GeV~} \lesssim M_{b} \lesssim 6\;10^{5}$~GeV, and $1\mbox{~meV~} \lesssim m_{b} \lesssim 1.5$~meV.

Measurements are already under way with a slightly modified version of the apparatus, where a Faraday-cell type polarization modulator has been inserted in the beam path to obtain simultaneous detection of rotations and ellipticities. During these measurements, Ne gas has been inserted at several pressures in the interaction region in order to change the optical path: this will allow a direct test of the supposed photon-boson oscillation phenomenon \cite{Venezia}.

Recently, limits on axion-photon coupling with a strength far from this estimate have been published in \cite{CAST}. While a detailed discussion of astrophysical bounds is beyond the scope of the present letter, we mention that recent theoretical work \cite{Masso:2005ym} suggests scenarios where it is possible to accommodate both results. 

Let us finally note here, that a crucial check of these speculations could come from a "photon-regeneration" experiment of the type already attempted in \cite{Cameron} and in \cite{Ruoso}.


{\it Acknowledgements} -- We wish to thank S. Marigo and A. Zanetti for their invaluable and precious technical help on the construction and running of the apparatus, and also R. Pengo and G. Petrucci who designed and built the rotating magnet system. We also acknowledge all the people of the Laboratori Nazionali di Legnaro that helped us and made this experiment possible.


\begin{thebibliography}{00}
\bibitem{Nota_1}
In this model, the induced rotation is caused by real production of bosons, while the induced ellipticity is due to virtual production.
\end{thebibliography}
\end{document}